\documentstyle[pre,aps]{revtex}
\begin{document}
\draft
\title{Free energy landscape of a dense hard-sphere system}
\author{Chandan Dasgupta\cite{jnc}}
\address{Department of Physics, Indian Institute of Science, Bangalore 
560012, India}
\author{Oriol T. Valls}
\address{School of Physics and Astronomy and Minnesota 
Supercomputer Institute, \\ University of Minnesota, 
Minneapolis, Minnesota 55455}
\date{\today}
\maketitle
\begin{abstract}
The topography of the free energy landscape in
phase space of a dense hard sphere system characterized by a discretized
free energy functional of the Ramakrishnan-Yussouff form is investigated
numerically using a ``microcanonical'' Monte Carlo procedure. 
We locate a considerable number of glassy local minima of the free
energy and analyze the distributions of the free energy at a minimum and
an appropriately defined phase-space ``distance'' between different minima.
We find evidence for the existence of pairs of closely related 
glassy minima (``two-level systems''). We also investigate the way the system
makes transitions as it moves from the basin of attraction of a minimum
to that of another one after a start under nonequilibrium conditions.
This allows us to determine the effective height of free energy
barriers that separate a glassy minimum from the others. The dependence
of the height of free energy barriers on the density is investigated in
detail. The general appearance of the free energy landscape resembles 
that of a putting green: relatively deep minima separated by
a fairly flat structure. We discuss the connection of our results with the 
Vogel-Fulcher law and relate our observations to other work on the
glass transition.
\end{abstract}

\section{Introduction}
\label{sec:intro}
When a liquid is cooled to temperatures below the equilibrium freezing
temperature at a sufficiently fast rate to prevent
crystallization, it enters a metastable supercooled state. As the
temperature is lowered further, the supercooled liquid undergoes a
glass transition to a state in which it behaves in most ways like
a disordered solid. The dynamic behavior of supercooled liquids near
the glass transition exhibits many peculiar features
\cite{rev,rev1,rev2}, such as multi-stage, non-exponential 
decay of fluctuations and a rapid growth of relaxation times,
which are not fully understood theoretically.

An intuitively appealing description that
is often used \cite{pwa,pgw} for qualitative explanations of the
observed behavior near the glass transition is based on the
so-called ``free energy landscape'' paradigm. The starting point of
this description is a free energy functional that expresses the free
energy of a liquid as a functional of the time-averaged local number
density. At high temperatures (or at low densities in systems, such as
those consisting of hard spheres, where the
density is the control parameter), this free energy functional is
believed to have only one minimum that represents the uniform liquid
state. As the temperature is decreased  to values near the equilibrium
crystallization temperature (or {\it mutatis mutandi} the
density is increased), a new minimum representing the crystalline
solid, characterized by a periodic modulation of the local density, 
should also develop. In the ``free energy landscape'' paradigm, it 
is assumed that a large number of ``glassy'' local minima
of the free energy, characterized by inhomogeneous but aperiodic
density distributions, also come into existence at temperatures close
to the equilibrium crystallization temperature. If the system gets 
trapped in one of these glassy local minima as it is cooled
rapidly from a high temperature, crystallization does not occur and
the subsequent dynamics of the system is governed by
thermally activated transitions among a subset of the large number of 
metastable glassy minima. If the system visits a large number of these minima
during its evolution over a relatively long observation time, 
it behaves like a liquid
over such time scales, in the sense that the time-averaged local density
remains uniform. However, the dynamic behavior in this regime, being
governed by thermally activated transitions over free energy barriers
of varying height, is expected to be slow and complex. In this picture,
the glass transition occurs when the time scale of transitions among 
the glassy minima becomes so long that the system remains confined in
a single ``valley'' of the landscape over experimentally accessible 
time scales.

The general features of the free energy landscape described above are
very similar to those found in analytic studies \cite{kw,kt,ktw} of
certain generalized spin glass models with infinite-range interactions.
Similar features have also been found in recent studies
\cite{bm,parisi} of spin
models with complicated infinite-range interactions, but no quenched
disorder. The equilibrium and dynamic behavior of these mean-field 
models exhibit a
striking similarity with the phenomenology of the glass transition.
These results suggest that the free energy landscape paradigm may
indeed provide a fitting framework for the development of a theoretical
understanding of the behavior of supercooled liquids near the glass
transition. The development of such a description would obviously
require detailed information about the topography of the free energy
landscape of dense supercooled liquids. Since the analytic
methods used in the aforementioned studies of mean-field models with 
infinite-range
interactions can not be readily generalized to study physical systems with 
short-range interactions, investigations of the properties of the free
energy landscape of simple glass-forming liquids require the use of 
appropriate numerical methods.

We have carried out a number of numerical studies
aimed at elucidating the relation between the dynamic behavior of
simple model liquids and the structure of the free energy surface in 
phase space. These studies, carried out for a dense hard-sphere system,
are based on a model free
energy functional proposed by Ramakrishnan and Yussouff(RY) \cite{RY}.
A discretized version of this free energy functional was found \cite{cd1,cd2}
to exhibit a large number of glassy local minima at densities close
to or higher than the value at which equilibrium crystallization 
occurs. [The control parameter for a hard-sphere system is the dimensionless 
density $n^* \equiv \rho_0 \sigma^3$, where $\rho_0$ is the average
number density in the fluid phase and $\sigma$ is the hard-sphere
diameter; increasing 
(decreasing) $n^*$ has the same effect as decreasing (increasing) 
the temperature of systems for which the temperature is the relevant
control parameter.] From numerical studies \cite{lvd,dv94,vd95}
of a set of Langevin equations appropriate for this system, we found that the 
nature of the dynamics changes qualitatively 
at a ``crossover'' density near $n^*_x = 0.95$. The dynamics of a
system initially prepared in the uniform liquid state continues to be
governed by small fluctuations near the uniform liquid minimum of the free
energy as long as the density is lower than this crossover value. 
For values of $n^*$ higher than the crossover density, the dynamic
behavior is governed by transitions among the glassy minima. 
The time scales for such transitions were estimated 
from a standard Monte Carlo (MC) method in Ref.\cite{dv96} and found to 
increase rapidly with increasing density. 

In this paper, we present the results of a numerical study in
which a new approach is used for further
investigations of the properties of the free energy landscape
of a dense hard-sphere system. This study is based on the
discretized free energy functional \cite{RY} used in our previous work. 
The development of an understanding of the dynamics of the system 
in the regime where it is governed by transitions among the glassy minima
of the free energy requires information about properties of the free 
energy landscape such as the number of glassy minima, the distribution
of their free
energies and overlaps, the heights of the saddle points that 
connect different glassy minima, and how the system evolves from 
one minimum to another through these saddle points. One also needs to 
determine the 
dependence of these quantities on the average density which, as mentioned
above, is the relevant control parameter for the hard-sphere system. 
In the present study, we have developed and
used a ``microcanonical'' MC procedure to obtain quantitative
information about some of these features of the free energy landscape. 
As described in section II below, this MC procedure enables us
to study in detail the process of transition between different glassy
minima of the free energy and thus provides valuable information about
the topography of the free energy surface in phase space. We have also
located a large number of glassy minima of the free energy in the
course of this study. This gives us useful information about 
some of the relevant statistical properties of the collection of glassy
minima and the dependence of these properties on the density. The main
results obtained from this study are summarized below.

By performing a study of the probability of transition from a 
particular glassy minimum to any other as a function of the free
energy increment (the excess free 
energy measured from that at the original minimum), we find that the
value of the free energy at which transitions to other minima begin 
to occur with a high probability is nearly the same for different 
glassy minima. This observation suggests that the
free energy surface in phase space has a ``putting green like'' 
topography in which the glassy minima are like ''holes'' of varying
depth embedded in a relatively flat background. The total number
of glassy minima is
a sensitive function of the discretization scale and the sample size.
For ``commensurate'' (as defined below)
values of these quantities, which allow the
existence of a crystalline minimum (which, when present, is the global 
minimum of the free energy at high densities), the number of glassy minima is
relatively large. Systems with incommensurate values of the
discretization scale and the sample size exhibit no crystalline minimum
and a substantially smaller number of glassy minima. For this reason, we have
carried out all our studies of the statistical properties of glassy
minima for a commensurate system. We find that the total number of
glassy minima for such a system remains nearly constant as the density
is varied in the range $0.94 \le n^* \le 1.06$. The free energies of the
glassy minima are distributed over a wide range between the free
energy of the uniform liquid and that of the crystalline solid. 
The width of this range increases as the density is increased. This
observation, together with the result that the number of minima is nearly
independent of the density, implies that the number of minima per unit
interval of the free energy (the ``density of states'' of glassy
minima) decreases with increasing density. A suitably defined ``phase 
space distance'' between
two different glassy minima also shows a broad distribution. Our study
shows the existence of pairs of glassy minima that differ
from each other in the rearrangement of a very small number of
particles. The height of the free energy barrier that separates two
minima belonging to such a pair is found to be quite small. Such pairs
may be identified as ``two-level systems'' which are believed \cite{tls}
to exist in all glassy systems. 

Our study of the probability  of transition from a particular glassy minimum
to the others as a function of the free energy increment and the MC
``time'' $t$ allows us to define an effective barrier height that
depends rather weakly on $t$. Some of our results for the dependence 
of this barrier height on the density have been briefly reported
\cite{dv98} in a recent paper.
As described there, we found that the growth of this 
effective barrier height with increasing density is consistent with a 
Vogel-Fulcher form \cite{vf} appropriate for a hard-sphere system
\cite{wa81}. From our numerical results about how the dependence of the 
effective barrier height on $t$ changes as the 
density is increased, we were able to conclude that the growth of 
the barrier height (and the consequent growth of the relaxation time)
is primarily due to entropic effects arising from an increase in the 
difficulty of finding low free-energy paths (saddle points) that 
connect one glassy local minimum with the others. Some of the details
not included in Ref.\cite{dv98} are provided in the present paper. We
also relate the new results described above with the conclusions
reached in Ref.\cite{dv98}.

The rest of this paper is organized as follows. In section II, we
define the model system studied and describe the 
numerical methods used. Section III contains a detailed
description of the results obtained in our study. Finally, in section
IV, we summarize the main conclusions and discuss them in
the context of other related work on the glass transition. 

\section{Model and Methods}
\label{section:m&m}

In this section we define the model free energy used in our study, and 
introduce the ``microcanonical'' MC method
that we have developed as a means of studying the topography of the
free energy surface of the model in phase space. 
We also discuss in detail the 
initial conditions and parameters used.

\subsection{The free energy}
\label{ssection:fe}

As explained in the Introduction, our system is 
characterized by a free energy functional $F[\rho]$ which is of the
following form \cite{RY}:

\begin{eqnarray}
F[\rho] &=& F_l(\rho_0)+ k_B T \left[ \int{d {\bf r}\{\rho({\bf r})
\ln (\rho({\bf r})/\rho_0)-\delta\rho({\bf r})\} } \right. \nonumber \\
&-& \left. (1/2)\int{d {\bf r} \int {d{\bf r}^\prime
C({|\bf r}-{\bf r^\prime|}) \delta \rho ({\bf r}) \delta
\rho({\bf r}^\prime)}} \right],
\label{ryfe}
\end{eqnarray}
where $F_l(\rho_0)$ is the free energy of the uniform liquid at density
$\rho_0$, and $\delta \rho ({\bf r})\equiv \rho({\bf r})-\rho_0$ is the
deviation of the density $\rho$ at point ${\bf r}$ from $\rho_0$.
We take our zero of the free energy at the uniform liquid value, i.e.
we set $F_l(\rho_0)$ equal to zero. In Eq.(\ref{ryfe}),
$T$ is the temperature and the function $C(r)$ the direct pair correlation
function \cite{hm86} of the uniform liquid at density $\rho_0$, which
we express in terms of the dimensionless density 
$n^*\equiv \rho_0 \sigma^3$ by making use of the Percus-Yevick
\cite{hm86} approximation. This approximation is known to be quite
accurate if the value of $\rho_0$ is not very high, and should
be adequate for all the densities ($n^* \le 1.06$) considered
in this study. It is well-known
\cite{hm86} that the direct pair correlation function of simple model
liquids characterized by an isotropic, short-range pair-potential with 
a strongly repulsive core (such as the Lennard-Jones liquid) is very
similar to that of the hard-sphere system at high densities. Therefore,
we expect the results obtained from this study to apply, at least
qualitatively, to other dense model liquids.

To perform the numerical calculations, we discretize our system
by introducing a three
dimensional cubic lattice of size $L^3$ and mesh constant
$h$ in which a discrete set of variables,
$\rho_i,\, i= 1, L^3$, are defined as
$\rho_i \equiv \rho({\bf r}_i) h^3$, where $\rho({\bf r}_i)$ is the
density at mesh point $i$.
It is often convenient, in performing and describing the
calculations, to deal with a dimensionless, normalized free energy 
per particle $f[\rho]$ defined as:

\begin{equation}
f[\rho]= \beta F[\rho]/N
\label{f}
\end{equation}
where $N = \rho_0 (Lh)^3 = n^* L^3 a^3$ is the total number of
particles in the simulation box, $\beta \equiv 1/(k_BT)$ and 
$a$ is the ratio $h/\sigma$.

\subsection{The microcanonical Monte Carlo method}
\label{ssection:micro}

Our main objective in this work is to find an efficient way to investigate
the topography of the free energy landscape of the hard-sphere system 
described by the discretized form of the free energy functional defined
in Eq.(\ref{ryfe}). Basically, what one would like to do is to start
the system in a known free energy state (e.g. a glassy local minimum
of the free energy), and then
investigate the topography of the free energy surface near the starting
point by allowing the system to evolve in time, finding out which
configurations it subsequently visits and where it ends up. 
A conventional Metropolis algorithm MC
procedure, as performed at lower densities in our
previous work \cite{dv96}, is not the most efficient way of doing 
this: From a computational point of view, a certain amount of computer time 
is spent at every step of a conventional MC simulation in
evaluating the exponential of the free energy change. More important,
in a conventional MC simulation carried out at the
rather high densities we will consider here, it would take a
very long time for the
system to move out of the basin of attraction of the minimum 
in which it is initially placed. This
makes a conventional MC study of the process of transitions among
free-energy minima  prohibitively expensive in the density
range we consider.

In order to overcome these difficulties of a standard MC simulation, 
we have devised another procedure which we call the ``microcanonical'' MC
method. The reason for the quotes surrounding the word
``microcanonical'' is that in our method the quantity
that is constrained to be lower than a specified value is not the
energy but the free energy $F$ defined in Eq.(\ref{ryfe}). 
Specifically, the procedure works as
follows: we choose a trial value of what we call the free energy increment,
which we denote by $\Delta F$, or alternatively by $\Delta f$ if we are 
dealing with the dimensionless version of Eq.(\ref{f}). Then, starting with
initial conditions which, as discussed below, correspond to a
configuration where the free energy is at a local minimum, we perform a MC
simulation in which we sweep the sites $i$ of the lattice sequentially.
At each step and site, we pick another site $j$ at random from the ones
that lie within a distance $\sigma$ from the site $i$. 
We then attempt to change the values
of $\rho_i$ and $\rho_j$ to $p(\rho_i+\rho_j)$ and $(1-p)(\rho_i+\rho_j)$,
where $p$ is a random number distributed uniformly in $[0,1]$. 
The attempted change
is accepted, and this is the crucial point, if and only if the free energy
after the change is less than $F_{max}\equiv F_0 +\Delta F$ where
$F_0$ is the initial value, that is, the value of the free energy at
the minimum where we start the computation.
The simulation proceeds up to a maximum ``time'', $t_{m}$,
measured in MC steps per site (MCS).

In implementing this procedure, they key point is that we 
perform a sweep over a range of values of $\Delta F$, with the same
initial conditions. Obviously, if $\Delta F$ is ``too small'', meaning
that it is smaller than the height of the lowest free energy barrier
between the starting minimum
and any other minima that may be ``nearby'', the system is going to
remain in the basin of attraction of the starting minimum. As we
increase $\Delta F$, there will eventually be 
one or more accessible minima, that is, minima
that the system can find within a ``time'' $t<t_m$. These minima
are separated from the initial minimum by free energy barriers of
height less than $\Delta F$.
The system may then move to a region of phase space in the basin
of attraction of this new minimum, or one of several newly
accessible minima. As $\Delta F$ is further increased, additional minima
will be made accessible, and furthermore, since additional paths will become
available between the initial minimum and the minima already accessible at
smaller values of $\Delta F$, these minima may be reached in fewer MC steps.
Clearly, if one obtains the information of near which minima the
system is, and how long it takes to get there, one can begin to map out
the free energy landscape.

To find out which basin of attraction the system is in at a 
certain time $t$, we save the configurations (i.e. the values of the 
variables $\rho_i$) at suitable, relatively
frequent, time intervals $\Delta t$. These configurations are then used
as the inputs in a minimization procedure \cite{cd1} that determines which
basin of attraction the system is in. This is done by finding the local
free energy
minimum the system moves to when one tries to minimize the free energy 
by making small changes in the variables $\rho_i$ in such a way that $F$
is always lowered. 

The entire procedure, that is, running the simulation up to a certain
time $t_m$ for a set of values of $\Delta F$, saving the configurations
at intervals $\Delta t$, and analyzing them, must obviously be repeated
a certain number of times (the ``number of runs'') and averaged over.
Furthermore, suitable ranges of values for $t_m$, $\Delta F$ and
$\Delta t$ must be fixed at the beginning from theoretical considerations
coupled to trial runs. Nevertheless, these steps would also be required
for a standard MC procedure and therefore the need for them does not
in any way detract from the efficiency of the microcanonical method.

We have carried out the numerical procedure outlined above at a number
of densities in the range $0.94 \le n^* \le 1.06$. We did not consider
densities lower than 0.94 because our earlier work \cite{dv94,vd95} has
shown that the dynamics of the system is governed by transitions among
glassy local minima only at higher densities. Since, as mentioned above,
the Percus-Yevick approximation used for the direct correlation
function $C(r)$ appearing in Eq.(\ref{ryfe}) becomes 
less accurate at relatively high densities
\cite{hm86}, values of $n^* > 1.06$ were not considered.

\subsection{Initial states and system parameters}
\label{ssection:specs}

Our computations were performed for two different sets of the two
computational system parameters -- the sample size $L$ and 
the mesh size $h$. In one case we took these two parameters to be
commensurate with a close-packed lattice, and in the other incommensurate.
This was done chiefly in order to study the dependence of
the structure of the free energy landscape on
the commensurability properties of the computational
system parameters, as well as on their values. We also
considered two different kinds of initial conditions, so that we could
investigate the topography of the free energy surface in different
regions of phase space.
The computationally more intensive part of our 
simulations was carried out for systems of size $L = 15$ 
with  periodic boundary
conditions and mesh size $h = \sigma/4.6$. These values of $L$ and
$\sigma$ are incommensurate with a close-packed lattice, and
as a result no crystalline
minimum was found for these samples. Two kinds of initial
conditions were used for such systems.
The first kind is the same as that used in Ref.\cite{dv96}.
These are configurations obtained by first allowing the system
to evolve from a uniform initial state under Langevin dynamics 
\cite{lvd,dv94} until its free energy (which, we recall, includes a
current dependent term in the Langevin model) reaches zero (indicating
the departure of the system from the basin of attraction of 
the uniform liquid minimum of the free energy), and then using the 
minimization procedure to reach
the minimum whose basin of attraction the system is in at that point.
That minimum configuration is then the starting point of the present
work. All the minima found this way exhibit glassy structure, as
determined by the form of the two-point correlation function (see
below) of the local density variables $\rho_i$. At higher densities, 
where the Langevin computation is
inappropriate, the minima found at lower densities were scaled up
by running the minimization program at the higher density using the
lower density configuration (which, of course, is not a minimum at the
higher density) as the starting point.

The other portion of the computations was performed for systems
with  $L = 12$
and $h = 0.25 \sigma$. These values are commensurate with a
close-packed (fcc) structure, so that a crystalline minimum is found at
sufficiently high densities.  Starting configurations used for 
simulations carried out for such samples were obtained by using the
minimization procedure discussed above on randomly inhomogeneous 
initial configurations. 
Out of several glassy minima found this way, we selected a few with
structures similar to that of the minima used in simulations of the $L
= 15$ sample. Because of the smaller size of these samples,  we were
able to explore more extensively several
aspects of the problem under consideration.

Our computations for the $L=15$ sample were carried out for a time
range $t_m=15000$ MCS. Computations for the system with $L=12$ were usually
carried out to $t_m=8000$ MCS. In both cases the
density range $0.94 \le n^* \le 1.06$ was covered. For the larger size and 
longer maximum time, an interval $\Delta t=5000$ MCS was used, while
a closer spacing, $\Delta t=2000$ MCS, was chosen for $L=12$. 

The structure of a local minimum of the free energy may be
characterized by the two-point correlation function $g(r)$ of the frozen
local density variables $\rho_i$ at the minimum. This function is
defined as
\begin{equation}
g(r) = \sum_{i>j} \rho_i \rho_j f_{ij}(r)/[\rho_{av}^2
\sum_{i>j}f_{ij}(r)], 
\label{gofr}
\end{equation}
where $\rho_{av} \equiv \sum_i \rho_i/L^3$ is the average value of the
$\rho_i$ at the minimum (values of $\rho_{av}$ vary from one glassy
minimum to another, but are always slightly higher than $\rho_0 h^3$,
the value of $\rho_i$ at the uniform liquid minimum), and $f_{ij}(r) =
1$ if the separation between
mesh points $i$ and $j$ lies between $r$ and $r+\Delta r$ ($\Delta r$
is a suitably chosen bin size), and $f_{ij}(r) = 0$ otherwise. In
Fig.\ \ref{fig1}, we have shown the pair correlation functions for two typical
minima used as initial states in our simulations. From the
structure of $g(r)$ shown in this figure, it is clear that both these 
minima are glassy. It is also apparent that the structure of the $L =
12$ minimum is quite similar to that of the $L = 15$ one. Other minima
used in our simulations have a similar structure.

\section{Results}
\label{section:r&d}
In this section, we describe in detail the numerical results obtained from our
study, and present our analysis of the numerical data. 

\subsection{Transitions between minima}
\label{ssection:trans}

First, we discuss the qualitative behavior of the system as it evolves
in ``time'' under the microcanonical MC ``dynamics'' from the initial
state at $t=0$ to $t=t_m$ as described in subsection
\ref{ssection:micro}. During the evolution
of the system, we monitor the dimensionless free energy $\beta F$ and
the maximum and minimum values
of the discretized density variables $\rho_i, i = 1,L^3$. The
maximum value is useful for detecting possible transitions to the
neighborhood of the uniform liquid minimum. If the system
fluctuates near one of the inhomogeneous minima of the free energy,
then the maximum value of $\rho_i$ would be much higher than the value
(close to $\rho_0 h^3$) it would have if the system were in the vicinity
of the uniform liquid minimum. We find that the system does not move to
the neighborhood of the liquid minimum for the values of $\Delta F$
considered here. The total free energy is found to remain nearly
constant at a value slightly lower than the maximum allowed value, 
$F_{max} = F_0 + \Delta F$. 

In some of the runs, we have also monitored, at much more frequent
time intervals, a
quantity $d(t)$ that measures the ``phase space distance'' of the
system point at time $t$ from the starting point at $t=0$. This
quantity is defined as
\begin{equation}
d^2(t) = \sum_i [\rho_i(t) - \rho_i(0)]^2,
\label{doft}
\end{equation}
where $\rho_i(0)$ are the values of the density variables at the
minimum from where the simulation is started. By monitoring the
time-dependence of this quantity, we obtain useful
information about how the system explores the free energy landscape
as it evolves in time. We find that if the value of the
free energy increment $\Delta F$ is 
small enough so that the system remains 
confined in the basin of attraction of the original minimum over the
duration of the simulation, then the phase space distance $d(t)$
saturates (or continues to increase very slowly) after a rapid 
initial increase. The value at which $d(t)$ levels off increases as
$\Delta F$ is increased. For values of $\Delta F$ that are sufficiently
large for the system to be able to
move to the basins of attraction of other
minima, the transitions to other basins of attraction are usually (but
not always) indicated by sudden increases in the value of $d(t)$.
Typical results for the time-dependence of $d^2(t)$ for three different values
of $\Delta F$ are shown in Fig.\ \ref{fig2}. The data 
shown were obtained for a $L
= 12$ system at $n^* = 1.02$. For $\Delta f =1.0$ and $\Delta f=1.4$,
the system was found to remain in the basin of attraction of the
initial minimum during the time scale (8000 MCS) of the simulation. In
the run with $\Delta f = 1.9$, the system was found to have moved to the 
basin of attraction of a different minimum at $t = 2000$ MCS. It moved to 
the basin of attraction of the crystalline minimum between times
$t=2000$ MCS and $t = 4000$ MCS, and stayed there for the 
remaining part of the run.
While any signature of the first transition from the initial 
minimum to the intermediate one is not clearly visible in the
time-dependence of $d(t)$ (possibly due to the overlap of any such
signature with the initial rapid increase of $d(t)$), the subsequent 
transition to the crystalline minimum is clearly indicated by a rapid 
rise (and eventual saturation) of $d(t)$. 

We now turn to our results for the process of transition of the system 
from the basin of attraction of the initial glassy minimum to the basins
of other minima of the free energy. The main quantity that we use to 
present our analysis of these results is 
the ``critical'' value of $\Delta f$, that is, the value of this
quantity at which the system begins to find 
other minima with a high probability. This value, which we will 
denote as $\Delta f_c(t)$ (or
$\Delta F_c(t)$ as the case may be), is algorithmically defined as follows:
at every time investigated (i.e. times 5000, 10000 and 15000 MCS 
for $L=15$
and times 2000, 4000, 6000 and 8000 MCS for the $L=12$ samples),
we test for progressively increasing
values of $\Delta f$ what is the probability,
$P(\Delta f,t)$, that the
system has moved to the basin of attraction of a free energy minimum
distinct from the one in which it was started. This probability, which 
we obtain by averaging over a sufficient number of runs, is 
(at constant time) obviously zero
for very small $\Delta f$ and rises toward unity as $\Delta f$ 
increases. At a constant $\Delta f$, it increases somewhat with 
MC time, as the system is allowed to explore further regions
of phase space. We define $\Delta f_c(t)$ as the value of $\Delta 
f$ for which, at that time, the switching probability reaches $1/2$. 
Note that the quantities $P$ and $\Delta f_c$ are also functions
of $n^*$. We find that the number of runs required to obtain 
$\Delta f_c$ reliably and
reproducibly is about ten to fifteen. The appropriate range of 
$\Delta f$ to be investigated is then determined by the need to hit
the middle range of the switching probability values. This requires
a bit of trial and error initially, but once the range is found,
then running simulations at values of $\Delta f$ successively stepped up by 
increments of $0.05$ is sufficient. 

The whole procedure is illustrated
in Fig.\ \ref{fig3} and Fig.\ \ref{fig4}, where we have 
shown the results for the 
transition probability as a function of the free energy increment
$\Delta f$ for a $L = 12$ minimum at $n^* = 1.02$ (Fig.\ \ref{fig3}), and a 
$L = 15$ minimum at $n^* = 0.99$ (Fig.\ \ref{fig4}). 
The validity of our procedure
for estimating the values of $\Delta f_c$ from the numerical data (the
estimated values of $\Delta f_c$ are indicated in these figures) 
can be judged from the plots. It is clear from the data shown in
these figures that the uncertainty in the estimated values of $\Delta
f_c$ is $\sim 0.05$, the spacing between successive values of $\Delta
f$ used in the simulation. 

As mentioned above, the determination of the probability of transition
as a function of $\Delta f$ requires repeating the numerical procedure
a number of times for a fixed set of values of $\Delta f$. We find that
the minima to which the system moves for values of $\Delta f$ close to
or higher than $\Delta f_c$ are, in general, different for
different runs. This is more obvious for $L=12$ samples which, as
discussed below, exhibit a larger number of distinct glassy minima.
This observation suggests that $\Delta f_c$ represents
a measure of the free energy increment for which a 
relatively large region of
phase space becomes accessible to the system. Another observation that
supports this interpretation is that the system almost never returns to
the basin of attraction of the initial minimum after making a
transition to the basin of attraction of a different one: after having
left the initial minimum, the system cannot find
its way back. In a few
runs, we found transitions at relatively small values of $\Delta f$
which are {\em always} to the basin of attraction of the same minimum.
In most of these cases, the new minimum was found to be 
very ``close'' in phase
space (as measured by the quantity defined in
Eq. (\ref{doft})), to the initial one. These are
examples of so-called ``two-level
systems'' discussed in more detail in the next subsection. In a few
cases, we found that the new minimum to which all the transitions occur
at low values of $\Delta f$ is not close to the initial one. These are
examples of ``special'' paths with low barrier heights which connect
the initial minimum with another specific minimum. Since such
transitions and the ones between minima which are very close to each
other do not correspond to the opening up of large regions of phase
space, we did not include such transitions in the calculation of the
transition probability. 

In this way, we determined $\Delta f_c$ as a function of the density
$n^*$ and Monte Carlo time $t$ for the ranges of time and
density mentioned above. Some of our results for $\Delta f_c$
are shown in Fig.\ \ref{fig5} for a $L = 12$ minimum and four 
different values of 
$n^*$, and in Fig.\ \ref{fig6} for a $L = 15$ minimum and also
four values of $n^*$. 
We find, as implied above, that $\Delta f_c$ is a
weak function of time, and also of course a stronger function of $n^*$.
The dependence of $\Delta f_c$ on $t$ for a fixed value of $n^*$
becomes more pronounced as $n^*$ is increased. These dependences were
studied in detail in Ref.\cite{dv98}, where analytic
fitted forms are given. Part of this analysis and some of the
conclusions drawn from it are summarized in subsection
\ref{ssection:vf} for completeness. 
  
Finally, we note that although our model and numerical procedure are
different from those used in most existing numerical studies of dense
liquids (such studies use conventional MC or molecular dynamics to
simulate the behavior of models defined by a microscopic Hamiltonian), 
some of the general features found in existing simulations (and also in
experiments) are reproduced in our work. We find that if the value of
$\Delta F$ is such that $\beta F_{max}$ exceeds an upper threshold,
then the system moves within a few hundred MC steps to the vicinity of
the uniform liquid minimum. The value of this upper threshold is found
to be close to $\beta F=5.0$. This is the 
microcanonical analog of the melting 
transition. As mentioned above, this threshold value is
not crossed in the simulations from which the results described here
were obtained. We also find that in runs with $\beta F_{max}$ close to,
but lower than the upper threshold, the system moves to the basin of
attraction of the crystalline minimum (for $L=12$)
with a high probability. This is
nothing but the process of annealing:  it is well-known from
experiments and simulations that crystallization may be induced by
heating a glassy system to a temperature close to (but lower than) its
melting temperature and then cooling it down.

\subsection{Properties of glassy minima}
\label{ssection:minima}

In the course of our computations, we have located many of the
glassy minima of the free energy. As mentioned
above, for the ``incommensurate'' $L = 15$ sample used in our work, the
number of minima we
have located at each density is not large. The total number of minima
found for this sample varies in the range of four to six,
with some tendency to higher values in the lower part
of the $n^*$ range considered here.
The ``commensurate'' $L = 12$ sample exhibits, as we
shall see below, a substantially larger
number of minima, one of which is crystalline (fcc). For this reason,
we consider chiefly the results obtained for $L = 12$, for
which we can produce significant statistics, in this subsection.

While studying the process of transitions among the minima, we carried
out a large number of minimization runs with many different initial
states. The total number of such runs is of the order of $10^3$ for
each of the values of $n^*$ studied. While our procedure does not
correspond to an exhaustive search for all the local minima of the system,
the fairly large number of initial states considered for each value of $n^*$
ensures that, for $L=12$ at least, we did locate a large fraction
of the local free energy minima of the system.
So, the statistical information obtained from our study can
be expected to be representative of the full collection of local minima.

The total number of local minima of the $L = 12$ system remains nearly
constant as the density is varied in the range $0.96 \le n^* \le 1.06$.
This number is close to 25. The numbers for different values of $n^*$
show small variations, but there is no clear systematic trend in the
dependence of this number on the density. In most cases, a 
minimum found at a particular density may be ``followed'' to higher or
lower densities by
using the values of $\rho_i$ at the minimum at the first density as
inputs to the minimization program at the new density. In a few cases,
we find that a minimum disappears as the density is increased or
lowered, but such occurrences are rare. From these observations, we
conclude that the total number of glassy minima does not exhibit any
strong dependence on the density. Our limited investigation of the
variation of the free energies of the glassy minima with density
suggests that the ordering of the free energies remains the same (i.e.
free energies of different minima do not cross) as the
density is changed.

The free energies of these minima are distributed in a band that lies
between the free energy of the uniform liquid (which, we
recall, is taken to be the
zero of the free energy scale) and that of the crystalline solid. The
width of this band increases with increasing $n^*$. Since the number of
minima is approximately independent of the density, this implies that
the ``density of states'' of the glassy minima decreases as $n^*$ is
increased. Specifically, let $p(\beta F) \delta $ be the probability of
finding a glassy minimum with dimensionless free energy between $\beta
F-\delta/2$ and $\beta F + \delta/2$. 
We have calculated this quantity from our
data at different values of $n^*$. Representative results at two
densities, $n^*=0.96$ and $n^* = 1.02$, are shown in Fig. \ref{fig7}. 
The values of $\delta$ used are 4.0 and 8.0 for $n^*=0.96$ and $n^* =
1.02$, respectively. While the
distributions for the two densities are qualitatively similar, the
range of $\beta F$ over which $p(\beta F)$ is nonzero is clearly wider
at the higher density. The consequent decrease in the values of
$p(\beta F)$ with increasing density is also clearly seen. Both
distributions show peaks near the upper end, and tails extending to
substantially lower values. However, the lowest free energy of the
glassy minima is substantially higher than the free energy of the
crystalline minimum (for the crystalline minimum, $\beta F = -102.4$ for
$n^* = 0.96$ and $\beta F = -167.4$ for $n^*=1.02$). If the probability of
finding the system in a glassy minimum is assumed to be proportional to
the Boltzmann factor $e^{-\beta F}$, then only those minima with free
energies lying near the lower end of the band would be relevant in
determining the equilibrium and dynamic properties of the system. Our
results indicate that the number of such ``relevant'' minima decreases
with increasing $n^*$.

In the present study, we find
certain correlations between the free
energy of a glassy minimum and its structure. Similar correlations were
also found and described in some detail in Ref.\cite{dv94}. 
Specifically, we find that minima with
lower free energies have more ``structure'' (as indicated by e.g. the
heights of the first and second peaks of the correlation function
$g(r)$ defined in Eq.(\ref{gofr})) and higher density that those with
lower free energies.

We have also studied how the distributions of the local density 
variables in two distinct glassy minima differ from one another. 
To do this, we need a measure 
of the difference between the distributions of $\rho_i$ in two
glassy minima. This measure should satisfy the requirement that it 
yield a zero value for the difference between two configurations if one
of them can be mapped to the other by a symmetry operation of the
computational mesh. The symmetries of the cubic mesh used in
our computation include the 48 symmetry operations of a simple cubic
lattice and all translations, taking into account
the periodic boundary conditions. The quantity 
$d_m(1,2)$ that we have used to measure the
difference in the density distributions at two minima
labeled ``1'' and ``2'' is defined as
follows: 
\begin{equation}
d_m(1,2) = \frac{1}{2} min\{{\bf R}\} \sum_i [\rho_i^{(1)} - 
\rho_{{\bf R}(i)}^{(2)}]^2,
\label{dm}
\end{equation}
where $\rho_i^{(1)}$ and $\rho_i^{(2)}$ are the discretized densities
at two minima, $\bf R$ represents one of the symmetry operations
mentioned above, ${\bf R}(i)$ is the mesh point to which mesh point $i$
is transformed under $\bf R$, and $min\{{\bf R}\}$ means that the $\bf
R$ that minimizes the quantity on the right is to be taken. Since the
variables $\rho_i$ in an inhomogeneous minimum are close to one at the
mesh points corresponding to the locations of the ``particles'', and close
to zero at the other mesh points, the quantity $d_m$ basically measures
the number of particles whose positions are different in the two minima
being compared. In Fig. \ref{fig8}, we display
in histogram form the results for the distribution of
$d_m$ at two values of the density. The two distributions are
qualitatively similar. Both are small at small values of
$d_m$ and exhibit peaks near $d_m = 15$, which
corresponds to about half of the total number of particles having
different locations in the two minima. From these results, we
conclude that most of the glassy minima are rather different
from one another. The arrangement of the particles in the glassy minima
is also very different from that in a crystalline minimum, as indicated
by the observation that the value of $d_m$ almost always lies above 15 
if one of the two minima being compared is glassy and the other one
is crystalline. 

The degree of similarity between two different minima may also be 
quantified in terms of their ``overlap''\cite{ktw}. For the discretized
system considered here, the dimensionless overlap $q(1,2)$ between two minima 
labeled ``1'' and ``2'' may be defined in the following way:
\begin{equation}
q(1,2) = \frac{1}{\rho_{av}L^3} max\{{\bf R}\} \sum_i 
[\rho_i^{(1)} - \rho_{av}] [\rho_{{\bf R}(i)}^{(2)}-\rho_{av}],
\label{qdef}
\end{equation}
where $\rho_{av}$ is the
average value of the $\rho_i$, which is assumed to be the same in the
two minima, and $max\{{\bf R}\}$ means that the $\bf R$ that maximizes 
the quantity on the right is to be taken. Using the aforementioned fact
that at the glassy minima found in the density range considered here,
the values of $\rho_i$ are close to one at a small number of mesh
points and close to zero at others, the following approximate relation
between $q$ and $d_m$ may be derived easily:
\begin{equation}
q(1,2) \simeq 1-d_m(1,2)/N -\rho_{av},
\label{qdrel}
\end{equation}
where $N \equiv \rho_{av} L^3$ is the total number of particles in the
simulation box. The observation that the distribution of $d_m$ has a
peak near $d_m = N/2$ then implies that the distribution of $q$ peaks
near the value 0.5.

The distributions shown in Fig.\ref{fig8} extend down to values of $d_m$ as
small as 2 or 3, indicating that there are a few pairs of 
glassy minima which are very similar to each other. For each value of 
$n^*$, we find a small number (3-5) of such pairs of minima. To take 
an example, for $n^* = 0.96$, we have
found two minima, with free energies $\beta F = -47.7$ and $-47.0$, for
which the value of $d_m$ is 2.4. A detailed examination of the density
distributions at these two minima reveals that the main difference
between their structures comes from small displacements of just two
particles. Of course, these displacements also produce small changes in
the values of $\rho_i$ at neighboring mesh points. We believe that 
these pairs of minima are examples of ``two-level systems'' whose existence
in glassy materials was postulated \cite{tls} many years ago in order to
account for some of the experimentally observed low-temperature
properties. The height of the free energy barrier that separates two
members of a two-level system is expected to be low. Our observations
are consistent with this expectation. For the pair of minima mentioned
above, we find that if we start the system from the minimum with $\beta
F_0 = -47.7$ and carry out our numerical procedure for finding
transitions to other minima, the system begins to show transitions to
the minimum with $\beta F_0 = -47.0$ as the value of $\Delta f$ is
increased above 0.7. For $0.7 \le \Delta f \le 1.4$, {\em all} the
transitions are to the other member of the two-level system.
Transitions to other minima begin to appear only for higher values of
$\Delta f$. (As noted above, we did not include transitions between the
members of a two-level system in our calculation of $\Delta f_c$.)

We have also looked at how the quantity $ F_c = F_0 +\Delta F_c$,
which measures the value of the total free energy at
which transitions to other minima begin to occur with a high
probability, varies from one minimum to another. As 
exemplified by Fig.\ref{fig9},
where we present the results for $\beta F_c$ for four minima at
$n^*=0.96$ and for three minima at $n^*=1.02$, the value of this quantity
is nearly constant for each value of $n^*$. While the values of $\beta
F_0$ vary over a range of about 15 at $n^*=0.96$, and over a range of
about 40 at $n^*=1.02$, the calculated values of $\beta F_c$ are
nearly the same (within the error bars) for the different minima 
at both densities. This observation suggest a 
``putting green like'' free energy landscape in which the local minima
are like ``holes'' of varying depth in a nearly flat background. This
structure also implies that there is a strong correlation between the
depth of a minimum and the height of the barriers that separate it from
the other minima: the barriers are higher for deeper minima. 

\subsection{The Vogel-Fulcher law and entropic effects}
\label{ssection:vf}

As first pointed out in Ref.\cite{dv98},
there is a direct connection between our results and the Vogel-Fulcher
law \cite{vf}. We summarize this connection here. The basic
point is that the results for $\Delta f_c$ as shown in 
Figs.\ \ref{fig5} and \ref{fig6} are consistent with the form:
\begin{equation}
\Delta f_c(n^*,t) = \frac{a(t)}{n^*_c-n^*} + b,
\label{vfeq}
\end{equation}
where $a(t)$ is a weak function of $t$, $b$ is a constant, and 
the ``critical'' density $n^*_c$ is found to be independent of $t$ within 
the accuracy of our results. 
In Fig.\ref{fig10}, we show the data for $\Delta f_c$ for a $L=15$ 
minimum at times 5000, 10000 and 15000 MCS, and also the best fits of
the data to the form of Eq.(\ref{vfeq}) with $b = 0$. 
The parameter values for the best fits are: $a=0.31$, $n^*_c=1.19$ 
for $t=5000$; 
$a=0.30$, $n^*_c=1.22$ for $t=10000$; $a=0.27$, $n^*_c=1.23$ for 
$t=15000$. The form of Eq.(\ref{vfeq}) leads at once 
to the Vogel-Fulcher law appropriate for a hard sphere system 
\cite{wa81} since the characteristic time should be proportional to 
the exponential of $\beta \Delta F_c$. The values of
$n^*_c$ obtained from the fits, particularly at later times,
are very close to the random close packing
density, $n^*_{rcp} \simeq 1.23$. This is in agreement with the results
of molecular dynamics simulations \cite{wa81}.
The $L=12$ data yield similar values of $n^*_c$,
but with $b \simeq 1.0$. 

The weak dependence of $\Delta f_c$ on $t$ was also analyzed in detail
in Ref.\cite{dv98} where it was found that this dependence for all values
of $n^*$ and all the minima studied is well-described by the form:
\begin{equation}
\Delta f_c(n^*, t) = c(n^*) t^{-\alpha} + \Delta f_0,
\label{fit2}
\end{equation}
with $\alpha$ in the range $0.24-0.40$, and $\Delta f_0$ nearly
independent of $n^*$. The coefficient $c(n^*)$ was found to increase
with increasing $n^*$. Fits of the data to the form of Eq.(\ref{fit2})
are shown in Fig. 2 of Ref.\cite{dv98}. [This form agrees with 
Eq.(\ref{vfeq}) if $a(t) \propto t^{-\alpha}$ and $c(n^*) \propto
1/(n^*_c-n^*)$; our data are consistent with these conditions.] This
result suggests a physical interpretation of the observed Vogel-Fulcher
behavior. The quantity $\Delta f_0$ (the value of $\Delta f_c$ in the
$t \rightarrow \infty$ limit) provides a measure of ($\beta/N$ times) 
the height of the lowest free energy barriers that must be crossed in 
order to reach some of the other local minima of the free energy from
the one under consideration. As discussed
in detail in Ref.\cite{dv98}, the coefficient $c(n^*)$ may be
interpreted as a measure of the difficulty of finding low free-energy
paths to other minima. The observation that $c(n^*)$ increases with
$n^*$ while $\Delta f_0$ is nearly independent of $n^*$ then implies
that the increase of the effective barrier height with increasing $n^*$
is primarily due to ``entropic'' effects associated with the difficulty
of finding low-lying saddle points that connect a minimum with the
others. 

Therefore, a picture emerges from our work
as to the origin of the Vogel-Fulcher divergence. As the system evolves
over longer and longer times, the probability that it will find paths
to other minima involving jumps over lower and lower free energy barriers 
increases. At early times, the system can only explore nearby paths and
must then jump over whatever barrier is available in that region. At
longer times, a wider region is explored and the chances of finding a
path with a lower barrier increase. What our argument shows is that 
the Vogel-Fulcher law follows from the fact that the difficulty of
finding such low-free-energy paths to other minima increases with 
increasing $n^*$.

\section{Summary and Discussion}
\label{s&d}

We have developed and used in this work a numerical method to study the
topography of the free energy surface of a dense hard-sphere system
characterized by a model free energy functional.  At the relatively
high densities considered in this study, this system exhibits a complex
free energy landscape characterized by the presence of many glassy
local minima.  The number of accessed glassy local minima is found to
depend strongly on the commensurability properties of the
discretization scale $h$ and the sample size $L$ used.  For fixed
values of these parameters, the number of minima is nearly independent
of the density in the range studied.  In the case where $L$ and $h$ are
commensurate, a crystalline minimum is found and the number of glassy
minima accessed is large enough to allow for statistical study.  The
free energy values at its minima are distributed over a broad band
whose width increases with increasing density. The phase-space distance
between different minima shows a broad distribution with a peak near
the high end. However, there are a few pairs of minima which are very
close to each other in phase space and are separated by low free-energy
barriers. These, we believe, are examples of ``two-level systems''
which are expected to be present in all glassy materials.  We have
found in all cases a strong correlation between the depth of a minimum
and the effective height of free energy barriers that separate it from
the other minima: deeper minima have higher barriers. The observed
density-dependence of the effective barrier height is consistent with
the Vogel-Fulcher law. Our results indicate that this Vogel-Fulcher
growth is primarily due to an increase in the difficulty of finding
low-free-energy paths to other minima as the density is increased. 

Our results have close connections with those of a number of
recent studies of the equilibrium and dynamic properties of dense
supercooled liquids. We first discuss the relation of our observations with 
spin-glass-like theories \cite{kt,ktw,gp1} of the structural 
glass transition. These theories are based on the similarity between the
phenomenology of the structural glass transition in 
so-called ``fragile'' \cite{rev2} liquids and the behavior found in a class
of generalized mean-field spin glass models \cite{kw,ktn} with 
infinite-range interactions, and also in certain mean-field spin 
models with complicated 
multispin interactions but no quenched disorder \cite{bm,parisi}. 
At high temperatures, the free energy of these mean-field models,
expressed as a function of the single-site magnetizations, 
exhibits only one minimum -- the 
``paramagnetic'' one at which all the site magnetizations are zero. As
the temperature is lowered, an exponentially large number of
non-trivial local minima come into existence at a
characteristic temperature $T_d$. A ``dynamic transition'', 
characterized by a breaking of ergodicity, occurs at this temperature. 
At this ``transition'', the system gets trapped in the basin of
attraction of one of the newly developed local minima and remains 
confined in this basin for all subsequent times because 
the free energy barriers between different
local minima diverge in the thermodynamic limit in these
models. This ``dynamic transition'' does not have any signature in
the equilibrium behavior of the system. A thermodynamic phase
transition occurs at a lower temperature $T_c$ at which the
configurational entropy associated with the exponentially large number
of free-energy minima becomes non-extensive. 

In the suggested analogy between these models and the structural glass
transition, the paramagnetic minimum of the free energy is identified
with the one that represents the uniform liquid, and the role of the 
non-trivial local minima of the free energy is played by the glassy
local minima of the liquid free energy.
The ``dynamic transition'' found at $T_d$ in the mean-field spin models
is thought to be smeared out in liquids. This is because the free energy
barriers between different minima are expected to remain finite in 
physical systems with finite-range interactions. It has been suggested
\cite{kt,ktw,gp1} that the temperature $T_d$ should be identified with the
``ideal glass transition''  temperature of mode-coupling theories
\cite{mct} of the dynamics of dense liquids. This temperature is
supposed to signal the onset of activated processes in the dynamics.
The temperature $T_c$ is interpreted as the ``Kauzmann temperature''
\cite{kauz} at which the difference in entropy between the supercooled 
liquid and the crystalline solid extrapolates to zero. The relaxation
time of the supercooled liquid is supposed to diverge at the same
temperature. Heuristic arguments that suggest that this divergence is
of the Vogel-Fulcher form have been proposed \cite{ktw,gp1}. These
arguments are based on an entropic mechanism associated with the
vanishing of the configurational entropy at $T_c$.

The behavior found in our numerical study is in qualitative agreement
with this scenario. We find a characteristic density (we
recall once more that the density plays the role of the temperature in the
hard-sphere system we consider) at which a large number of glassy
minima of the free energy come into existence. We do not yet
know whether the number of glassy minima depends exponentially on the
sample volume -- a study of this question is difficult due to the
dependence of the number of minima on the commensurability of
$h$ and $L$. While the number of glassy minima for fixed $h$ and $L$ 
remains nearly constant as the
density is increased, the configurational entropy associated with these
minima decreases with increasing density because the width of the band
over which the free energy of these minima is distributed increases
with density. As discussed above, we have also found evidence for a
Vogel-Fulcher-type growth of relaxation times driven by an entropic 
mechanism.

There are, however, a number of differences between the details of our
findings and the predictions of spin-glass-like theories. In our
earlier work \cite{dv94,vd95} on the Langevin dynamics of the model
system considered here, we found that the dynamic behavior is governed by
activated processes if the dimensionless density $n^*$ exceeds a
crossover value, $n^*_x$, of about 0.95. This value is substantially higher than
the value of $n^*$ ($\approx$ 0.8) at which the glassy minima
make their first appearance. These two densities are assumed to be the same
in spin-glass-like theories. Another difference lies in the values of
the free energy of the glassy minima relative to that of the uniform
liquid. We find that the free energy of a glassy minimum becomes
lower than that of the uniform liquid minimum as the density is
increased above a value that is only slightly higher than the density at
which the glassy minimum comes into existence. In particular, the free
energies of the glassy minima are substantially lower than that of the
uniform liquid one for values of $n^*$ near $n^*_x$. This is 
different from the behavior found in the spin
glass models. In these systems, the free energies of the non-trivial
local minima remain higher than that of the paramagnetic one over the
entire temperature range $T_c < T < T_d$. Our results for the
distribution of the overlap between different minima are also somewhat
different from those for the spin glass models. We cannot rule out that
some of these differences arise from finite-size effects which may be
significant for the rather small samples considered in our 
study. Another possibility is that these differences arise in our
system from the effects of small fluctuations about a local minimum, 
which are unimportant in models with infinite-range interactions.
A careful investigation of these issues would be very interesting.

A number of numerical studies of ``aging'' phenomena in the
non-equilibrium dynamics of simple model liquids have
been reported recently \cite{kb1,kb2,gp2}. In these studies, the
system is quenched from a relatively high temperature to a temperature
lower than the numerically determined glass transition temperature, and 
is then allowed to evolve at this low temperature for a certain
``waiting time'' $t_w$. Then, the two-time correlation function
$C(t_w,t_w+t)$ of an appropriate fluctuating quantity is measured and
the dependence of this correlation function on $t$ and $t_w$ is
analyzed. The simulations show that the decay of $C(t_w,t_w+t)$ as a
function of $t$ becomes slower as $t_w$ is increased. Our results about
the topography of the free energy landscape provide a qualitative
explanation of this observation. When the system is rapidly quenched to a
low temperature (or compressed to a high density in our hard-sphere
system), it is likely to get trapped in the basin of attraction of one 
of the glassy local minima that are close in phase space to the initial 
configuration. Such minima would not, in general, have the lowest 
free energies. As the system evolves during 
the waiting time $t_w$, it can be expected to move progressively 
to the basins of
attraction of minima with lower free energies because such minima would
have a higher Boltzmann weight. Since the effective
barrier height is higher for deeper minima (this follows from the
``putting green like'' structure of the free energy landscape), the time
scale for subsequent relaxation is expected to increase with 
increasing $t_w$. This is precisely the behavior found in the aging 
simulations mentioned above.

Our study is rather similar in spirit to numerical investigations of the
``potential energy landscape'' \cite{sw83,ws85,s95,ah97,sds98,aprv,dpvr} of
model liquids characterized by simple Hamiltonians. In such studies, a
numerical minimization procedure (e.g. the conjugate gradient
method) is used to find the local minima of the total potential
energy of small samples as a function of the coordinates of the
particles. The potential energy function is generally found to exhibit 
a large number of local minima. These local minima and the 
potential energy barriers that separate them define a complex 
``potential energy landscape''. Recently, there have been
several attempts \cite{s95,sds98,aprv} to relate the properties of this
landscape to the dynamic behavior of the liquid. While the
similarities between these investigations and our work are obvious, 
there are several important differences
between these two approaches, some of which we now discuss.
A study of the potential energy landscape is based on a microscopic
Hamiltonian defined in terms of the coordinates of the particles,
whereas our work involves a model free energy which is a functional of
a coarse grained (both in space and time) density field. Information
about the microscopic interactions is incorporated in our description
through the direct pair correlation function $C(r)$ appearing in
Eq.(\ref{ryfe}). The free energy of a thermal system is, of course,
equal to the potential energy at zero temperature. Therefore, the
potential energy landscape of such systems becomes identical to the
free energy landscape at $T=0$. There are some mean-field spin-glass
models (e.g. the $p$-spin spherical spin glass \cite{pspin}) in which
the correspondence between the local minima of the energy and the free
energy extends also to non-zero temperatures.  Such a
correspondence is not likely to be generic, however. A description based on
the potential energy landscape is certainly appropriate at low
temperatures where entropic effects are relatively unimportant.
But it would, in general, be difficult to extend such a
description to higher temperatures where entropic effects play a
crucial role. In particular, information about the energy landscape
alone would not be sufficient to describe the behavior near a phase
transition (such as the melting transition of a solid and the
order-disorder transition in magnetic systems) driven by a competition
between energetic and entropic effects. In contrast, a description
based on a model free energy that includes entropic contributions
provides a convenient and intuitively appealing starting point for
studying the behavior near such phase transitions. For example, the free
energy functional used in our work is known \cite{RY} to provide a
correct description of the crystallization transition of simple
liquids. Another well-known example is the Curie-Weiss theory of
magnetism (our approach is analogous to an inhomogeneous version of the
Curie-Weiss theory). For these reasons, we believe that our
free-energy-based approach is more suitable for a description of the
behavior of liquids near the glass transition than approaches based on
the potential energy landscape.

Another important difference between free energy and potential energy
landscapes is that the former changes as the appropriate control
parameter (density or temperature) is changed, whereas the latter, being
determined completely by the Hamiltonian of the system, remains
unchanged. Specifically, some of the local minima of the free energy
may appear or disappear as the control parameter is varied (for
example, the inhomogeneous minima of the free energy used in our study
disappear at sufficiently high densities). Also, the heights of free
energy barriers between different local minima may change with the
control parameter (see, e.g. Fig.\ref{fig10} where the dependence
of the height of a typical free energy barrier on the density is
shown). In contrast, the potential energy landscape does not show 
any such variation as the temperature is changed. This
difference may be important in understanding some of the results found 
in recent studies \cite{aprv,dpvr} based on the energy landscape of
simple model liquids. In Ref.\cite{aprv}, an approximate description of
the dynamics of a Lennard-Jones system in supercooled and glassy
regimes is developed in terms of the numerically determined properties
of the local minima of the potential energy and the energy barriers
between them. While this description is found to reproduce several
interesting features of glassy dynamics, it {\em does not} show the
expected faster than Arrhenius growth of the viscosity at low
temperatures. This may be due to the energy barriers in
this description not changing with temperature. It is possible that a
free-energy-based description in which the barrier heights change
with temperature would lead to a faster than Arrhenius growth 
of the viscosity. This possibility is clearly illustrated in our study
which shows that the dependence of the heights of free energy barriers 
on the appropriate control parameter leads to Vogel-Fulcher behavior.
Ref.\cite{dpvr} describes a numerical study of the local minima and the
saddle points of the potential energy surface of small Lennard-Jones
clusters. One of the quantities calculated in this paper is an 
``entropic ratio'' $R$ that approximately quantifies the entropic effects on
the rate of thermally activated transitions between two local minima of 
the potential energy function. Values of $R > 1$ indicate entropic
suppression of the transition rate, whereas $R < 1$ corresponds to an
enhancement. The probability of $R$ having values greater than one is
found to be small. This result is interpreted as evidence for
entropic effects being relatively unimportant. In particular, the
authors mention that this observation contradicts our conclusion 
(described in detail in Ref.\cite{dv98} and summarized in section 
\ref{ssection:vf} above)
that the growth of relaxation times in simple glassy liquids is
primarily entropic in origin. In our opinion, the results reported in 
Ref.\cite{dpvr} do not necessarily contradict our conclusion. The
difference between our conclusion and that of Ref.\cite{dpvr} about the
importance of entropic effects is probably just a reflection of the
aforementioned fact that the free energy landscape changes with the
control parameter, but the potential energy landscape does not. We
find in our free-energy-based study that the {\em growth} of the 
height of a typical effective {\em free
energy} barrier with increasing density is primarily due to entropic
effects arising from an increase in the difficulty of finding
low-free-energy paths to other minima. This effect is closely
related to {\em changes} in the topography of the free energy landscape
as the density is changed. In contrast, the potential energy landscape 
studied in Ref.\cite{dpvr} {\em does not} depend on the temperature 
which is the appropriate control parameter for the Lennard-Jones
system considered there. In particular, the calculated values of the
height of the potential energy barrier between two minima and the 
entropic factor $R$ {\em do not} change as the temperature is changed. 
Therefore, there is no direct connection between our results (which, as
explained above, are about the {\em changes} of these quantities as 
the appropriate control parameter is changed) and those reported 
in Ref.\cite{dpvr}.

We end with a word of caution. Due to the computational complexity of
numerical studies of free energy and potential energy landscapes, such
studies have been restricted to rather small samples which may exhibit
finite-size effects. In our work, we found that certain features of the
free energy landscape are quite sensitive to the 
commensurability properties of the
discretization scale and the sample size. Strong dependences on the
sample size and the boundary condition have also been found
\cite{ah97,dpvr} in studies of the potential energy landscape. One
should, therefore, be careful in extrapolating the results obtained
from such studies to the thermodynamic limit. 

\section{Acknowledgments}
A part of the numerical work was carried out at the Supercomputer 
Education and Research Centre of Indian Institute of Science. One of us
(CD) acknowledges support from the Theoretical Physics Institute,
University of Minnesota, for a visit.

\begin{figure}
\caption{The density correlation function $g(r)$, as
defined in Eq.(\protect\ref{gofr}), plotted as a function
of distance (in hard sphere units) for two typical initial
configurations. Note the glassy character of the
correlations for both lattice sizes. }
\label{fig1}
\end{figure}
\begin{figure}
\caption{The quantity $d^2(t)$, which characterizes
the ``phase space distance'' between two points, (see
Eq.(\protect\ref{doft})) plotted as a function of Monte
Carlo time for three values
of the free energy increment. The regions of sharp
changes in the curves are discussed in the text.}
\label{fig2}
\end{figure}
\begin{figure}
\caption{Example of the determination of the 
``critical'' value $\Delta f_c$, defined as the value of the free
energy increment $\Delta f$ at which the transition probability $P$ is 1/2.
The black dots mark the intersections of the plots with
the line $P=0.5$ (see text for a complete discussion). The data shown
are for a sample of size $L = 12$.}
\label{fig3}
\end{figure}
\begin{figure}
\caption{Another example of the determination
of $\Delta f_c$ (see the caption of Fig.\protect\ref{fig3} for the details),
but for a sample of size $L=15$.}
\label{fig4}
\end{figure} 
\begin{figure}
\caption{Results for the critical value, $\Delta f_c$,
of the free energy increment, as a function of the Monte Carlo time $t$
for four different values of the dimensionless density $n^*$.
The results shown are for a $L=12$ minimum.}
\label{fig5}
\end{figure}
\begin{figure}
\caption{Results for $\Delta f_c$ as a function of the Monte Carlo time
$t$ at four different densities for a $L = 15$ minimum. In this figure
and in Fig.\protect\ref{fig5}, results for other times and
densities studied interpolate smoothly with the
results shown.}
\label{fig6}
\end{figure}
\begin{figure}
\caption{The ``density of states'' for glassy free energy minima, defined
as the probability of finding a glassy minimum with free energy in
a given range (see text). Results for $L = 12$ samples at two densities
are shown.}
\label{fig7}
\end{figure}
\begin{figure}
\caption{Histogram representing the fraction of pairs of free
energy minima found to differ in their real space
density configurations by an amount $d_m$ as defined
in Eq.(\protect\ref{dm}). Results for $L = 12$ samples are shown 
at two different densities.}
\label{fig8}
\end{figure}
\begin{figure}
\caption{The value $F_c$ of the free energy at which
transitions to other minima begin to occur with a high probability,
plotted as a function of $F_0$, the free energy at the starting minimum.
Results are shown for $L = 12$ at two densities. 
One can see that, at a given
$n^*$, the dependence of $F_c$ on the starting value is quite weak.}
\label{fig9}
\end{figure}
\begin{figure}
\caption{Vogel-Fulcher fits of the data for $\Delta f_c$ obtained for a
$L=15$ minimum at three different values (5000, 10000 and 15000
MCS) of $t$. The continuous lines are the best fits of the data to the form
of Eq.(\protect\ref{vfeq}) with $b=0$. The parameter values for the best fits
are given in the text.}
\label{fig10}
\end{figure}


\begin{references}
\bibitem[*]{jnc} Also at the Condensed Matter Theory Unit, Jawaharlal 
Nehru Centre for Advanced Scientific Research, Bangalore 560064, India.
\bibitem{rev} For a review, see {\it Liquids, Freezing and the Glass 
Transition}, edited by J. P. Hansen, D. Levesque, and J. Zinn-Justin,
(Elsevier, New York, 1991).
\bibitem{rev1} J. J\"{a}ckle, Rep. Prog. Phys. {\bf 49}, 171 (1986).
\bibitem{rev2} C. A. Angell, J. Phys. Chem. Solids, {\bf 49}, 863
(1988). 
\bibitem{pwa} P. W. Anderson in {\it Ill Condensed Matter, Lecture
Notes of the Les Houches Summer School, 1978}, edited by R. Balian, R.
Maynard and G. Toulouse (North Holland, Amsterdam, 1979). 
\bibitem{pgw} P. G. Wolynes in {\it Proceedings of International
Symposium on Frontiers in Science, (AIP Conf. Proc. No. 180)}, edited by
S. S. Chen and P. G. Debrunner (AIP, New York, 1988).
\bibitem{kw} T. R. Kirkpatrick and P. G. Wolynes, Phys. Rev. A
{\bf 35}, 3072 (1987); Phys. Rev B {\bf 36}, 8552 (1987).
\bibitem{kt} T. R. Kirkpatrick and D. Thirumalai, J. Phys. A
{\bf 22}, L149 (1989).
\bibitem{ktw} T. R. Kirkpatrick, D. Thirumalai and P. G. Wolynes,
Phys. Rev. A {\bf 40}, 1045 (1989).
\bibitem{bm} J. P. Bouchaud and M. Mezard, J. Phys. I (France)
{\bf 4}, 1109 (1994).
\bibitem{parisi} L. F. Cugliandolo, J. Kurchan, G. Parisi and F.
Ritort, Phys. Rev. Lett. {\bf 74}, 1012 (1995).
\bibitem{RY} T. V. Ramakrishnan and M. Yussouff, Phys. Rev. B {\bf 19},
2775, (1979).
\bibitem{cd1} C. Dasgupta, Europhys. Lett. {\bf 20}, 131 (1992).
\bibitem{cd2} C. Dasgupta and S. Ramaswamy, Physica A {\bf 186}, 314
(1992). 
\bibitem{lvd} L. M. Lust, O. T. Valls, and C. Dasgupta, Phys. Rev. E {\bf 48},
1787 (1993).
\bibitem{dv94} C. Dasgupta and O. T. Valls, Phys Rev. E {\bf 50}, 3916
(1994).
\bibitem{vd95} O. T. Valls and C. Dasgupta,  Transport Theory
and Stat. Physics {\bf 24}, 1199 (1995).
\bibitem{dv96} C. Dasgupta and O. T. Valls, Phys. Rev. E {\bf 53}, 2603
(1996).
\bibitem{tls} P. W. Anderson, B. I. Halperin and C. M. Varma, Philos. Mag.
{\bf 25}, 1 (1972); W. A. Phillips, J. Low. Temp. Phys. {\bf 7}, 351
(1972). 
\bibitem{dv98} C. Dasgupta and O. T. Valls, Phys. Rev. E {\bf 58}, 801
(1998).
\bibitem{vf} H. Vogel, Z. Phys. {\bf 22}, 645 (1921); G. S. Fulcher
J. Amer. Ceram. Soc. {\bf 8}, 339 (1925).
\bibitem{wa81} L. V. Woodcock and C. A. Angell, Phys. Rev. Lett. {\bf 47},
1129 (1981).
\bibitem{hm86} J. P. Hansen and I. R. McDonald, {\it Theory of Simple 
Liquids} (Academic, London, 1986).
\bibitem{gp1} G. Parisi in {\it Complex Behaviour of Glassy Systems:
Proceedings of the XIV Sitges Conference}, edited by M. Rubi and C.
Perez-Vicente (Springer, Berlin, 1997).
\bibitem{ktn} T. R. Kirkpatrick and D. Thirumalai, Phys. Rev. B {\bf
36}, 5388 (1987).
\bibitem{mct} See, for example, W. G\"{o}tze in Ref.\cite{rev}.
\bibitem{kauz} W. Kauzmann, Chem. Rev. {\bf 48}, 219 (1948).
\bibitem{kb1} W. Kob and J.-L. Barrat, Phys. Rev. Lett. {\bf 78}, 4581
(1997). 
\bibitem{kb2} J.-L. Barrat and W. Kob, 1998 preprint
(cond-mat/9806027). 
\bibitem{gp2} G. Parisi, J. Phys. A {\bf 30}, L765 (1997); Phys. Rev.
Lett {\bf 79}, 3660 (1997).
\bibitem{sw83} F. H. Stillinger and T. A. Weber, Phys. Rev. A {\bf 28},
2408 (1983).
\bibitem{ws85} T. A. Weber and F. H. Stillinger, Phys. Rev. B {\bf 32},
5402 (1985).
\bibitem{s95} F. H. Stillinger, Science {\bf 267}, 1935 (1995).
\bibitem{ah97} A. Heuer, Phys. Rev. Lett. {\bf 78}, 4051 (1997).
\bibitem{sds98} S. Sastry, P. G. Debenedetti, and F. H. Stillinger,
Nature {\bf 393}, 554 (1998).
\bibitem{aprv} L. Angelani, G. Parisi, G. Ruocco, and G. Villani, 1998
preprint (cond-mat/9803165).
\bibitem{dpvr} G. Daldoss, O. Pilla, G. Villani, and G. Ruocco, 1998
preprint (cond-mat/9804113).
\bibitem{pspin} J. Kurchan, G. Parisi, and M. A. Virasoro, J. Phys. 
I (France) {\bf 3}, 1819 (1993).
\end{references}
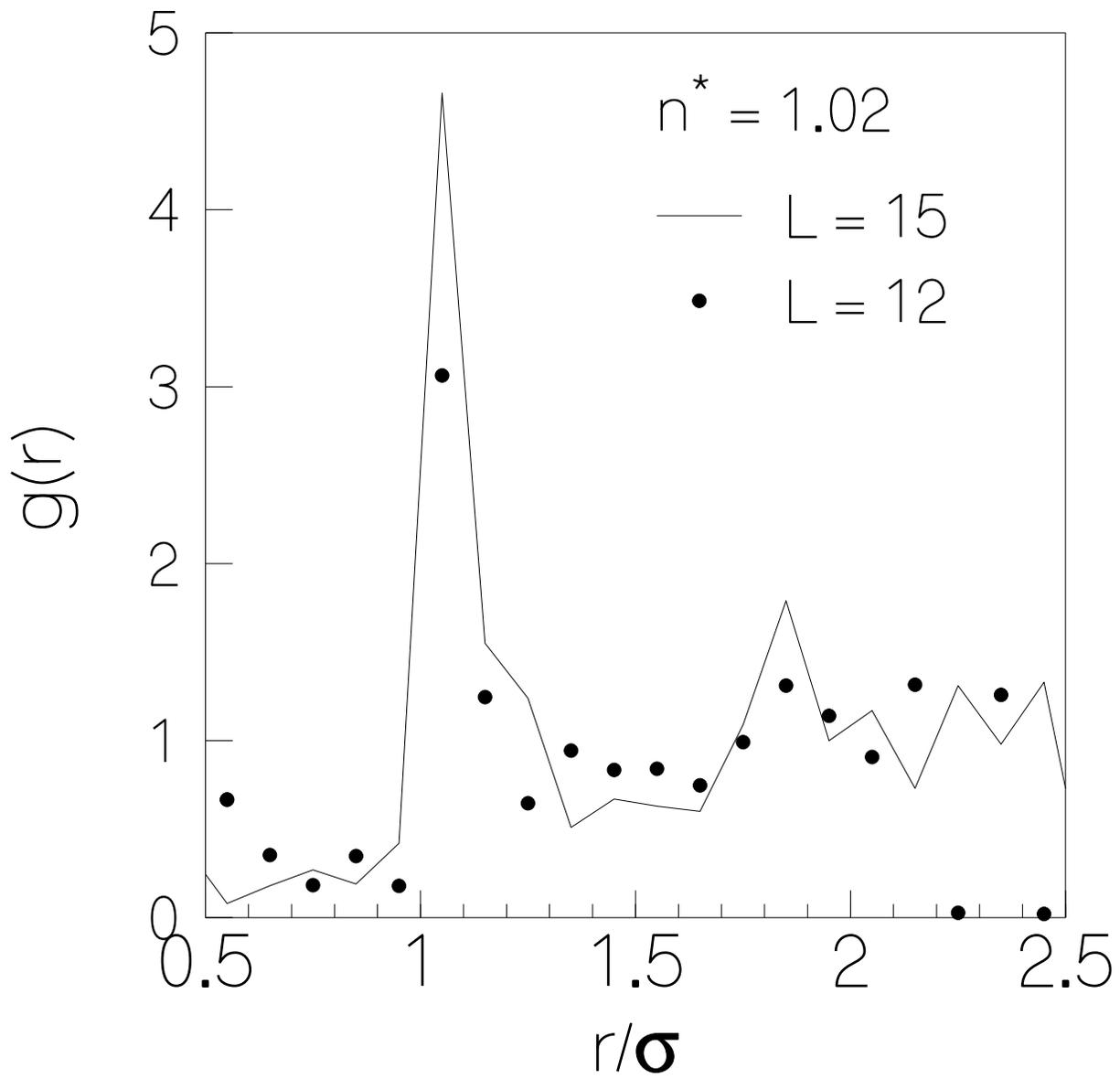
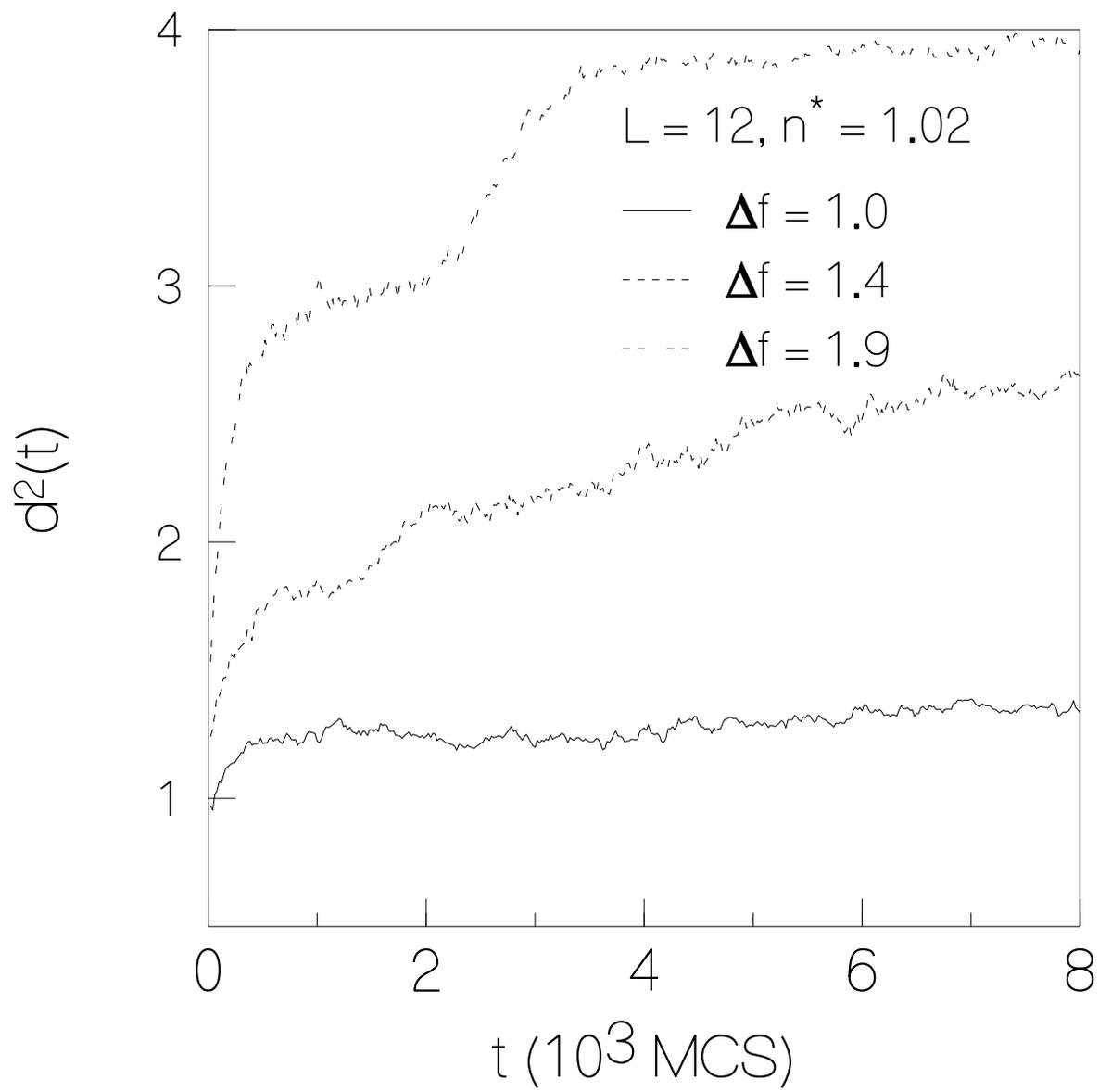
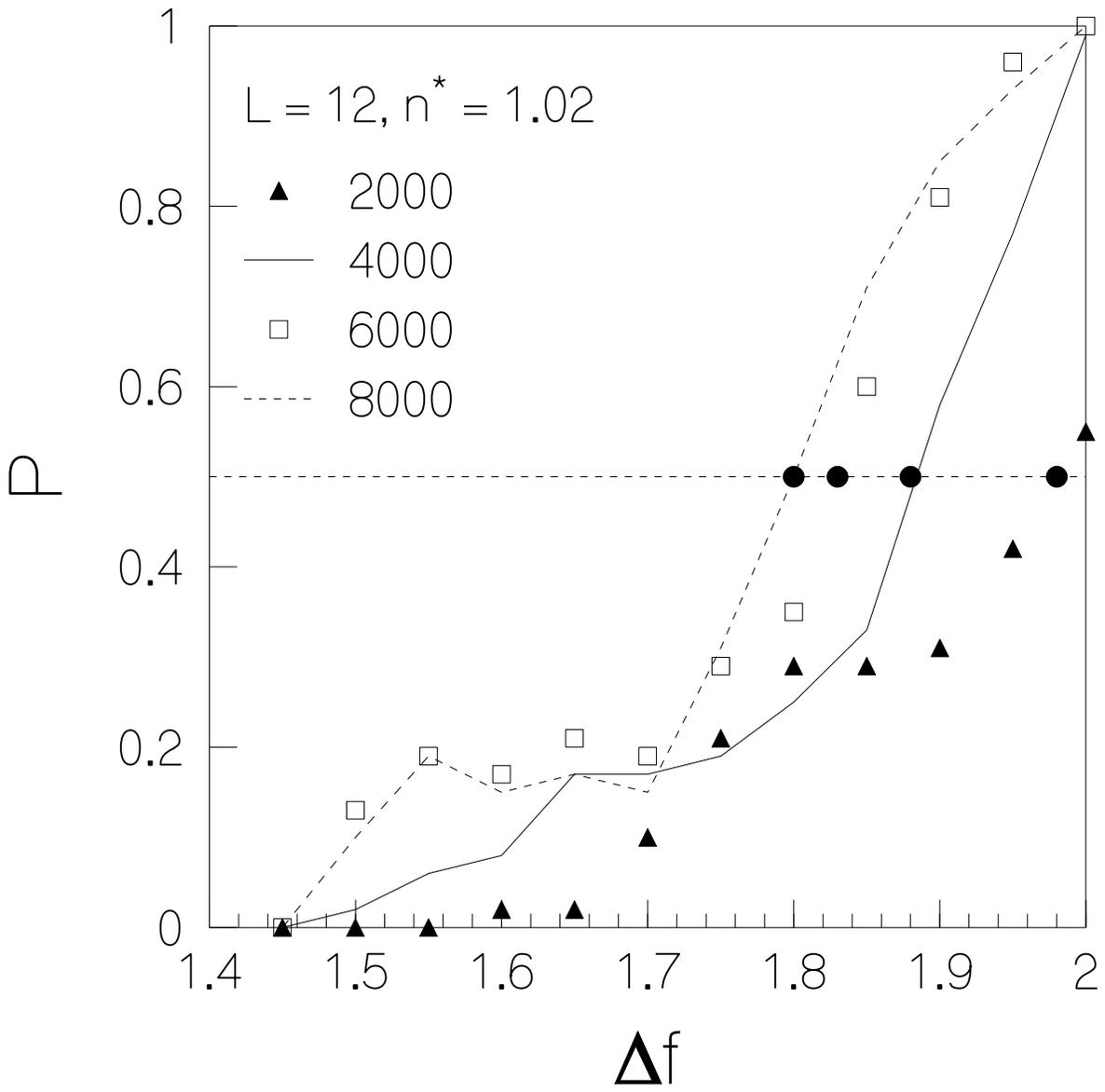
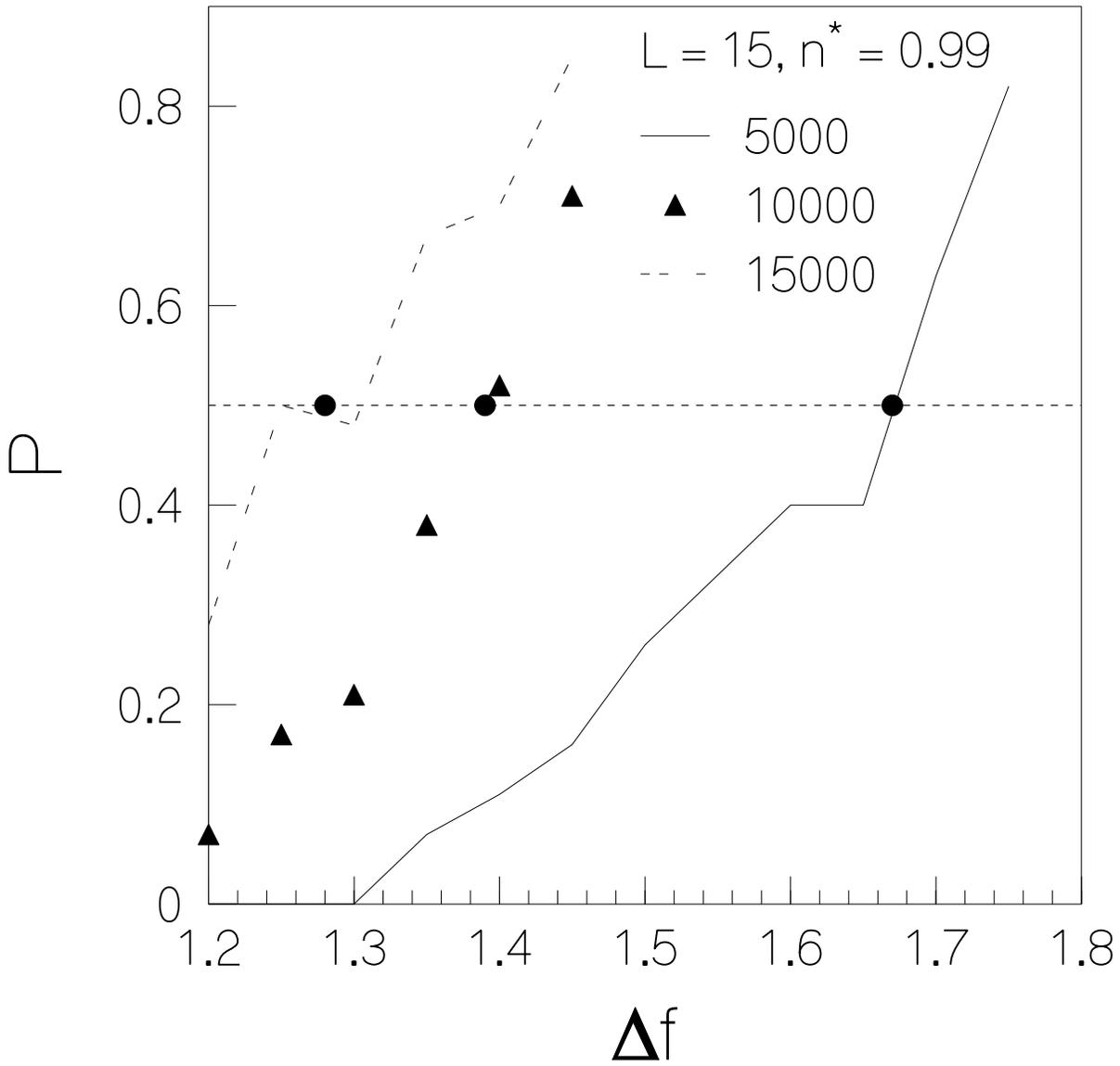
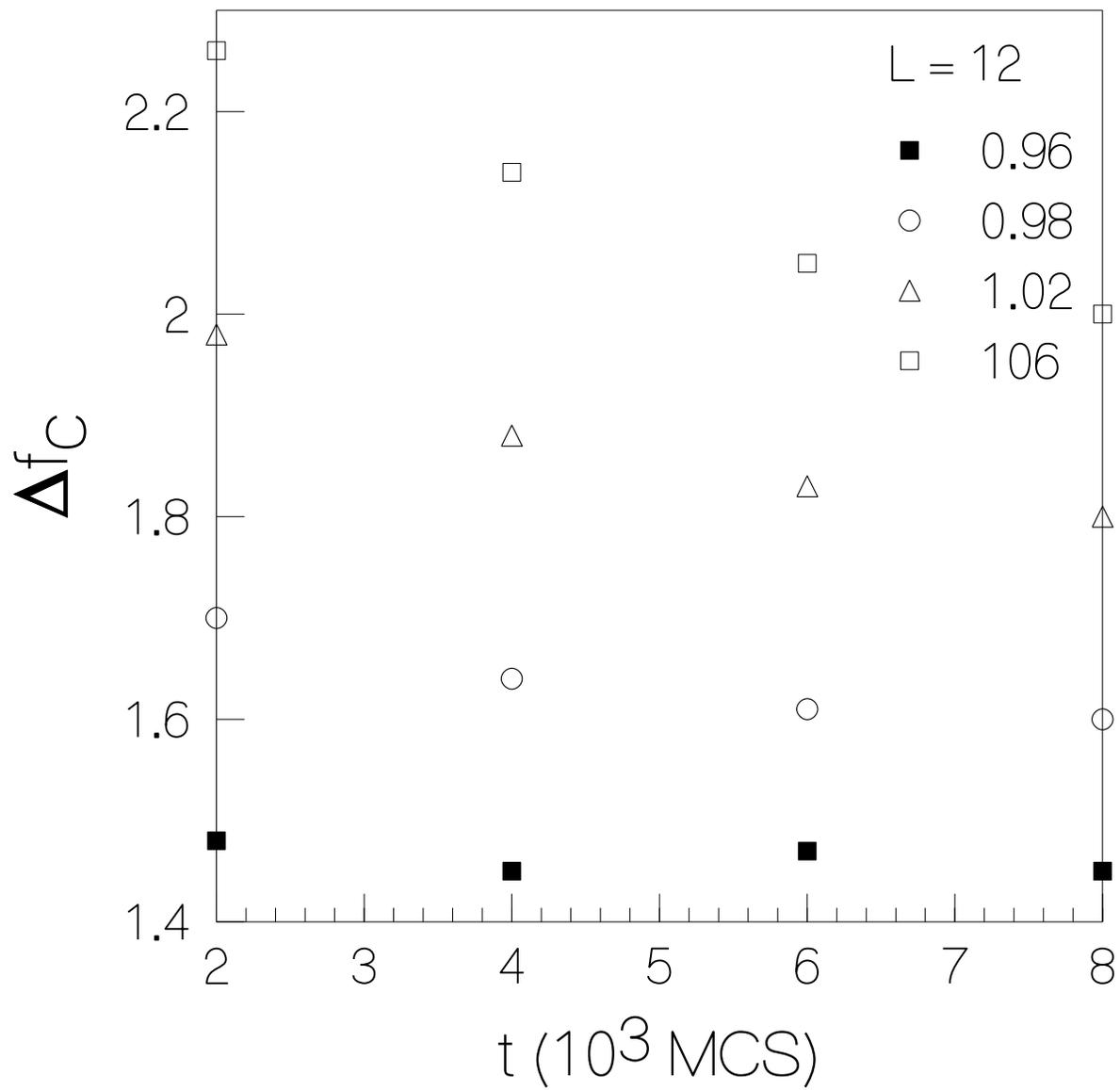
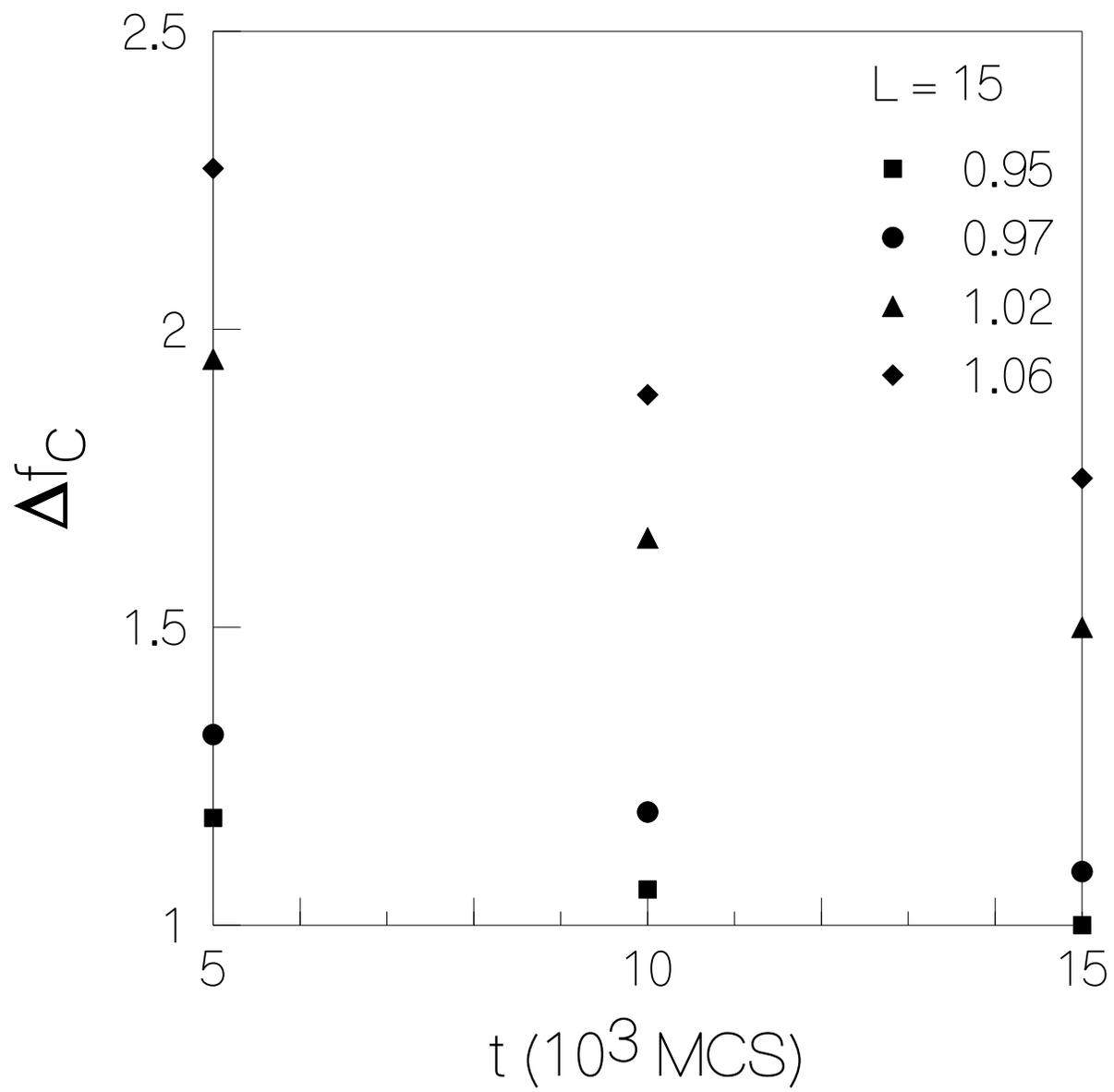
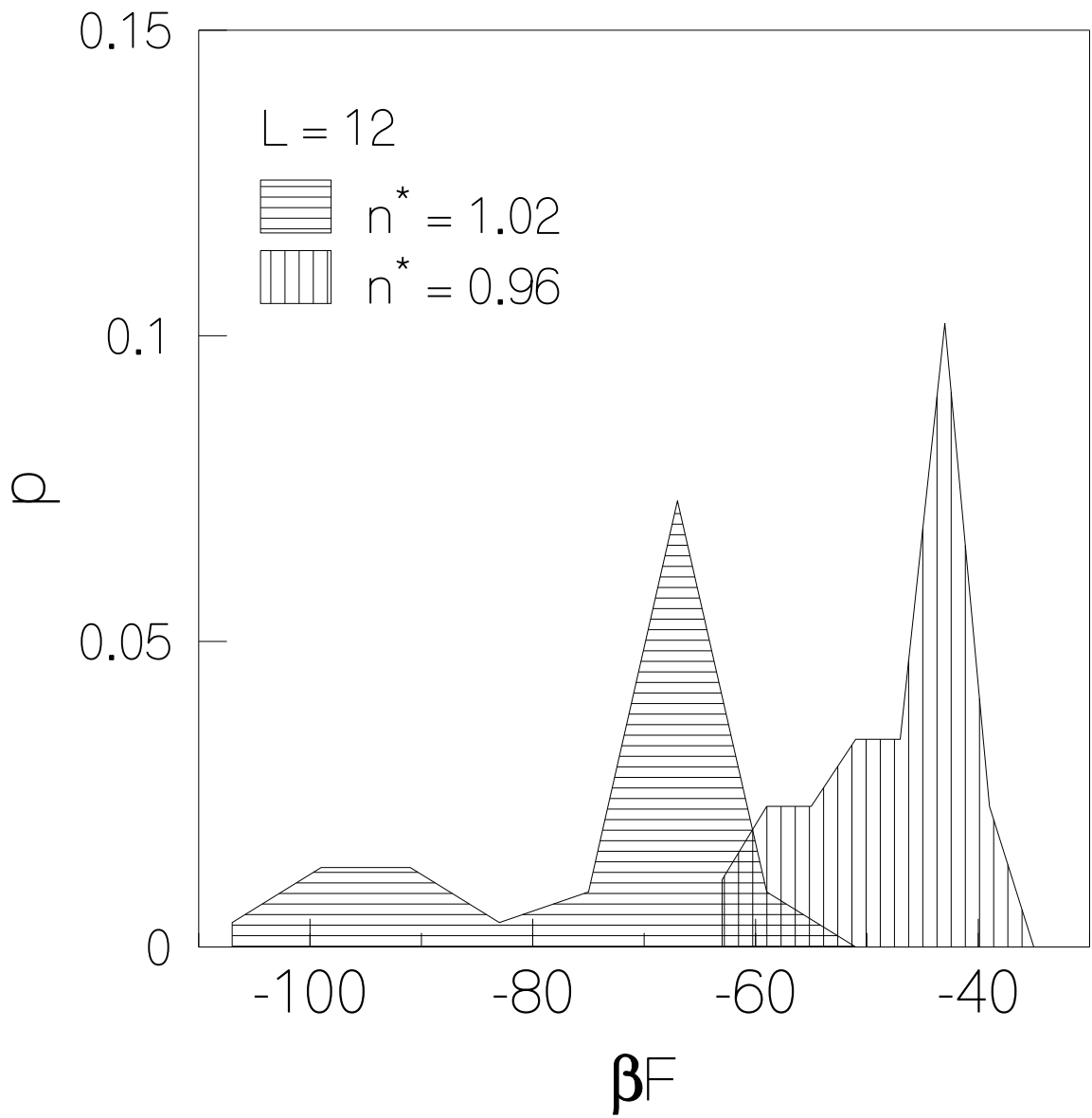
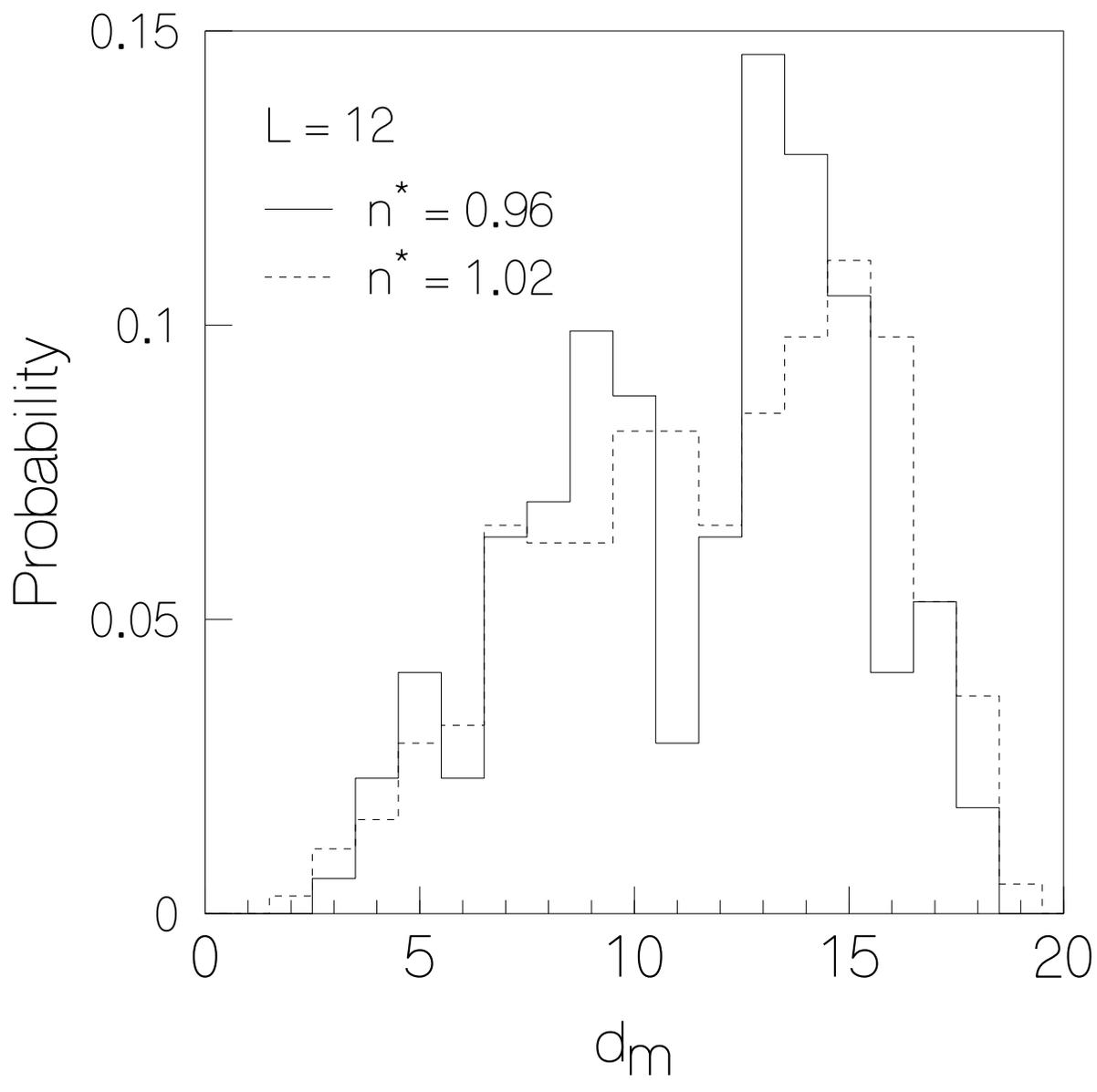
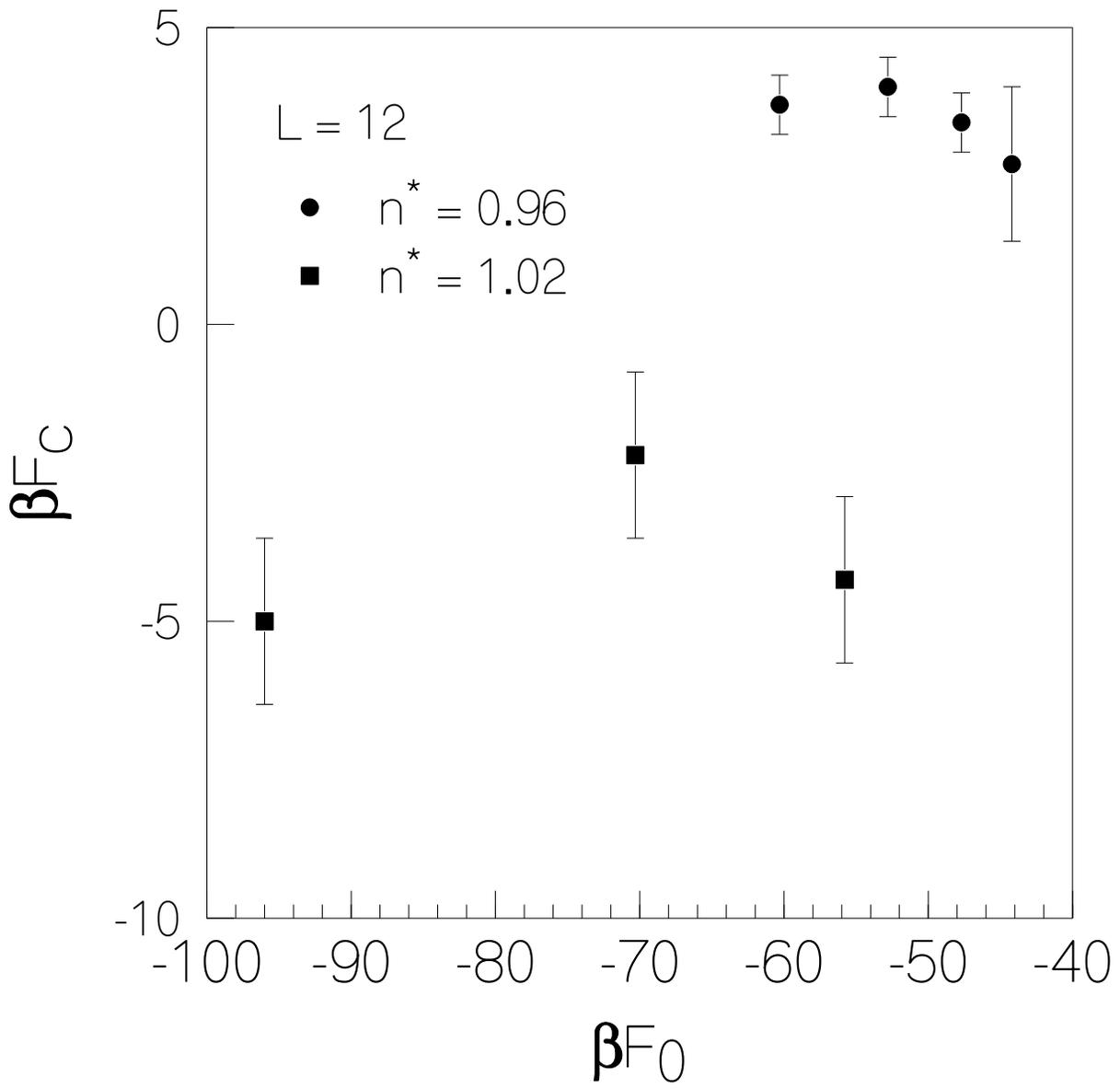
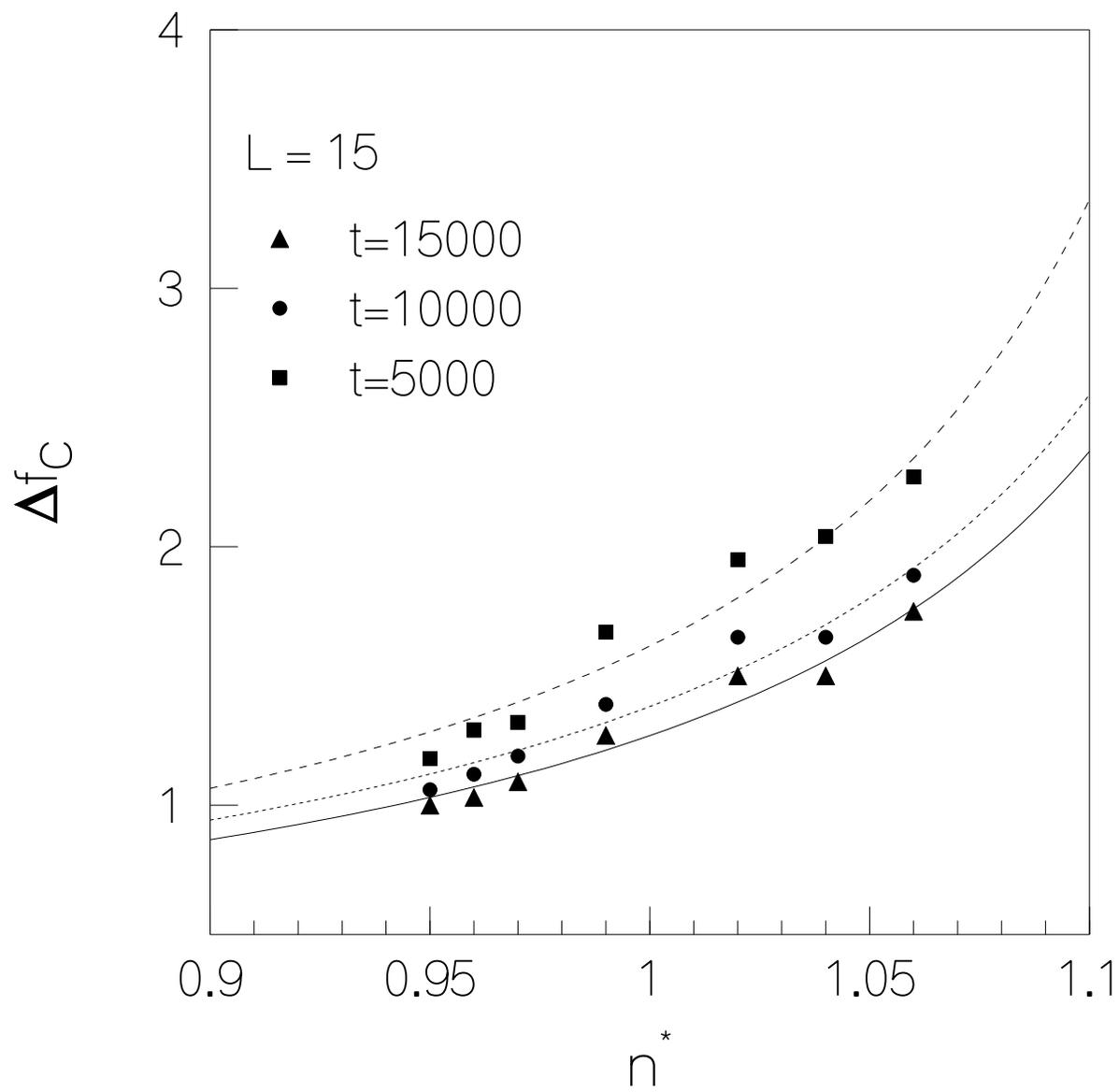
\end{document}